\documentclass{article}
\usepackage{graphicx,authblk,float, amsmath,authblk,graphicx,lipsum, fancyhdr, wrapfig,lineno}

\usepackage[english]{babel}

\usepackage[a4paper,margin=1in]{geometry}

\usepackage[sorting=none, style=nature]{biblatex} 

\usepackage[colorlinks=true, allcolors=blue]{hyperref}

\title{Physics-Informed ML Exploration of Structure-Transport Relationships in Hard Carbon}

\author[1,2,*]{Nikhil Rampal}
\author[1,2]{Stephen E. Weitzner}
\author[3]{Fredrick Omenya}
\author[1,2]{Marissa Wood}
\author[3]{David M. Reed}
\author[3]{Xiaolin Li}
\author[1]{Jonathan R. I. Lee}
\author[1,2,*]{Liwen F. Wan}

\affil[1]{Materials Science Division, Lawrence Livermore National Laboratory, Livermore, California, 94550, United States}
\affil[2]{Laboratory for Energy Applications for the Future (LEAF), Lawrence Livermore National Laboratory, Livermore, California, 94550, United States}
\affil[3]{Pacific Northwest National Laboratory, Richland, Washington, 99352, United States}

\affil[*]{rampal1@llnl.gov, wan6@llnl.gov}

\date{}

\begin{document}

\maketitle

\renewcommand{\abstractname}{\large Abstract}

\thispagestyle{fancy}

\begin{abstract}

\normalsize 

{Sodium-ion batteries are a cost-effective and sustainable alternative to lithium-ion systems for large-scale energy storage. Hard carbon (HC) anodes, composed of disordered graphitic and amorphous domains, offer high capacity but exhibit complex, poorly understood ion transport behavior. In particular, the relationship between local microstructure and sodium mobility remains unresolved, hindering rational performance optimization. Here, we introduce a data-driven framework that combines machine-learned interatomic potentials with molecular dynamics simulations to systematically investigate sodium diffusion across a broad range of carbon densities and sodium loadings. By computing per-ion structural descriptors, we identify the microscopic factors that govern ion transport. Unsupervised learning uncovers distinct diffusion modes, including hopping, clustering, and void trapping, while supervised analysis highlights tortuosity and Na–Na coordination as primary determinants of mobility. Correlation mapping further connects these transport regimes to processing variables such as bulk density and sodium content. This physics-informed approach establishes quantitative structure–transport relationships that capture the heterogeneity of disordered carbon. Our findings deliver mechanistic insights into sodium-ion dynamics and provide actionable design principles for engineering high-performance HC anodes in next-generation battery systems.
}

\hspace{1cm}

\end{abstract}


\newpage

\section*{Main}


Sodium-ion batteries (SIBs) are emerging as cost-effective and scalable alternatives for sustainable energy storage. Their appeal is driven by the abundance of sodium and the availability of low-cost materials, making SIBs particularly suitable for stationary applications where affordability and long cycle life outweigh the need for high energy density~\cite{sun2022understanding}. Among potential anode materials, hard carbons stand out for their high capacity, robust chemical stability, and tunable microstructure, which enables further performance optimization~\cite{fan2024research}.

Despite their promise, designing high-performance hard carbon anodes remains challenging due to an incomplete understanding of sodium storage mechanisms and the lack of predictive structure–property relationships. Hard carbons exhibit considerable structural complexity, featuring both short-range ordered (graphitic) regions and disordered or porous domains. Several mechanistic models have been proposed, starting with the classical “insertion–filling” model, which attributes the high-voltage sloping region to sodium intercalation into graphitic-like interlayers and the low-voltage plateau to pore filling within disordered domains~\cite{chen2021hard, chu2023reconfiguring, irisarri2015hard}. Subsequent frameworks, such as “adsorption–insertion” and “adsorption–filling,” have considered additional factors like graphitic stacking faults, edge sites, and distinctions between closed and open nanopores~\cite{sun2022understanding, chu2023reconfiguring, xie2020hard}. These models aim to explain sodium storage across the diverse microstructural domains of hard carbon, including graphitic layers, pores, and amorphous boundaries, rather than in uniformly amorphous systems.

Recent studies underscore that the effectiveness of sodium storage mechanisms in hard carbon is closely tied to its underlying microstructure. Hard carbon is inherently heterogeneous, displaying hierarchical motifs such as nanocrystalline graphitic domains, stacking faults, amorphous boundary regions, and both open and closed pores. These features collectively influence the balance between sloping and plateau capacities, affect ion transport kinetics, and ultimately shape electrochemical performance. For instance, graphitic interlayers serve as intercalation sites responsible for the sloping region of the voltage profile, while closed pores facilitate quasi-metallic sodium adsorption, contributing to the low-voltage plateau~\cite{chen2021hard, xiao2025insight, chu2023reconfiguring}. The size, distribution, and connectivity of these domains are highly sensitive to synthesis parameters, including pyrolysis temperature, precursor chemistry, and densification processes~\cite{fan2024research, sun2022understanding, zhang2016correlation}.

Importantly, these microstructural domains can be deliberately tuned: increasing graphitic content can enhance rate capability, while controlled pore formation can boost reversible capacity. However, while the roles of ordered domains are relatively well understood, the contributions of disordered and interfacial regions remain unclear, despite mounting evidence that they are critical for Na-ion mobility, clustering, and safety in practical electrodes.

The more disordered and amorphous domains, particularly those at the interfaces between crystalline-like regions, nanoscale surfaces, and pore walls, are structurally elusive but electrochemically significant, especially in high-capacity and high-rate applications. Their spatial and temporal heterogeneity complicates direct characterization, yet they can play a dominant role in sodium-ion mobility and clustering behavior. Recent operando and spectroscopic studies have highlighted the importance of these features. For example, Xiao \emph{et al}. demonstrated that closed pores with diameters around 1.6 nm offer an optimal balance between minimizing diffusion barriers and achieving favorable sodiation potentials, thereby reducing the risk of sodium metal plating at high currents~\cite{xiao2025insight}. Iglesias \emph{et al}. used operando X-ray techniques to show that such microstructural environments promote quasi-metallic sodium clustering within confined domains, reinforcing the link between pore curvature and electrochemical behavior~\cite{kitsu2025microstructure}. In situ NMR and SAXS analyses further corroborate these findings, confirming the progressive filling of closed nanopores and associating higher curvature with reduced sodiation potentials~\cite{fan2024research}.

From a mechanistic perspective, Sun \emph{et al}. and others have shown that defects and heteroatom doping can modulate adsorption energetics in the sloping region, although these effects are difficult to quantify due to the complex and amorphous nature of local environments~\cite{sun2022understanding, chu2023reconfiguring}. Computational studies have sought to address this challenge. Density functional theory (DFT) has yielded valuable insights into sodium binding at edge sites, vacancies, and carbon fragments, but remains limited to small-scale, idealized systems~\cite{li2022towards}. Reddy \emph{et al}. combined in situ Raman spectroscopy with first-principles calculations to propose a multi-stage sodium insertion pathway involving interlayer intercalation, surface adsorption, and pore filling~\cite{anji2018insight}. However, simulating realistic disorder and dynamic ion transport at scale continues to be a significant challenge. Reactive force fields such as ReaxFF allow for larger-scale simulations, yet their accuracy in capturing many-body interactions and ion clustering is limited~\cite{li2022insight}. Recent safety-focused studies further highlight the importance of understanding ion behavior in disordered domains. For example, Nio \emph{et al}. reported that sodium clustering within closed nanopores can create electronically metallic domains, lower thermal stability below 100\,\textdegree{}C, and mimic the effects of metal plating~\cite{niu2025sodium}.

Collectively, these observations implicate structural features such as tortuosity, pore morphology, and coordination environment as key determinants of both kinetics and safety. Despite recent advances, much of the current understanding still relies on ensemble-averaged observables and indirect ex situ analyses, which obscure local heterogeneity. A central knowledge gap remains: how do local atomic-scale features and structural disorder influence sodium mobility and phase behavior across the diverse environments present in real hard carbon materials?

To address this gap, we develop a machine learning-based interatomic potential trained on high-fidelity DFT data. Unlike traditional force fields, machine learning potentials can generalize across a wide range of local environments and capture complex many-body interactions, providing both the accuracy of quantum methods and the scalability needed for realistic simulations. This approach enables accurate and efficient large-scale molecular dynamics simulations that capture structural disorder, heterogeneity, and the dynamic evolution of sodium within highly disordered carbon matrices. By overcoming the fidelity–scale trade-off inherent in earlier methods, our framework provides simultaneous access to atomistic detail and long-timescale transport behavior, directly addressing the need to resolve local heterogeneity in sodium-ion dynamics.

Our integrated framework advances the field in three key ways. First, we extract trajectory-resolved structural descriptors, including ion coordination, accessible volume, and tortuosity. Second, we apply unsupervised learning techniques, such as agglomerative clustering, to identify distinct sodium-ion transport behaviors within the evolving carbon network. Third, we conduct correlation and feature importance analyses to determine the dominant structural features governing ionic diffusion. This physics-informed, data-driven approach reveals how sodium ions respond to their local environments, insights that are inaccessible to traditional, average-based analyses.

Ultimately, our work provides a predictive and interpretable understanding of sodium transport in hard carbon anodes. It establishes a robust processing–structure–property framework that links synthetic variables (such as densification and annealing) to microscopic structure and macroscopic performance. These insights offer practical guidance for designing next-generation carbon architectures that combine high-rate capability with thermal safety.

\section*{Results}

\subsubsection*{Physically-Guided Workflow for Diffusion Mode Classification}

\begin{figure}[H]
    \centering
    \includegraphics[width=\textwidth]{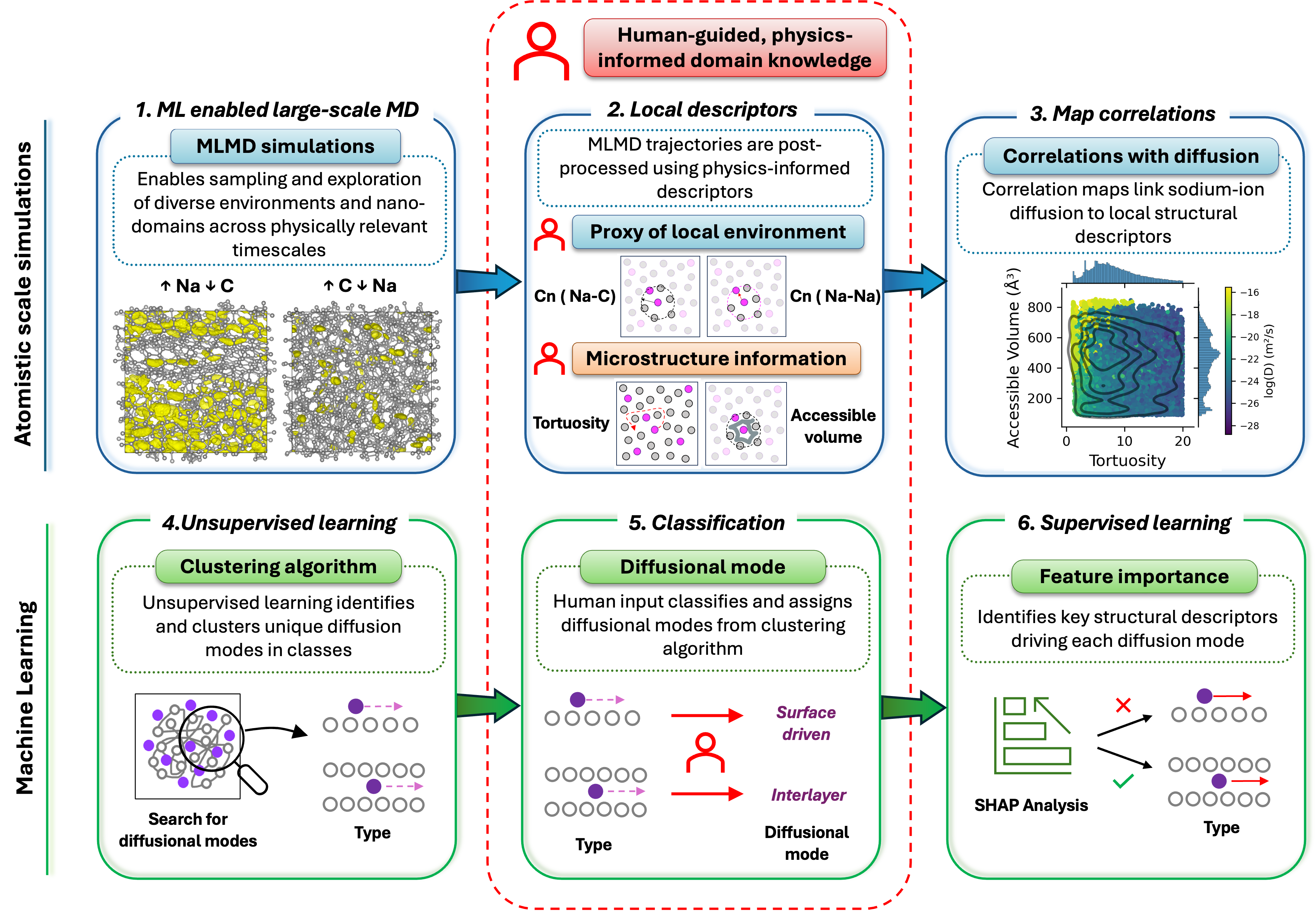}
    \caption{
    \textbf{End-to-end framework for sodium-ion diffusion analysis in disordered carbon using MLMD and human-guided, physics-informed machine learning.} 
    The workflow integrates machine learning molecular dynamics (MLMD) with both unsupervised and supervised learning to decipher ion transport mechanisms. 
    (\textbf{1}) Large-scale MLMD simulations are performed using machine-learned interatomic potentials (MLIPs), enabling exploration of sodium trajectories across long time and length scales. 
    (\textbf{2}) Human-guided, physics-based local descriptors are computed for each Na-ion, including Na--Na coordination, Na--C coordination, tortuosity, and accessible volume. 
    (\textbf{3}) Correlation maps link diffusion coefficients with structural descriptors, revealing structure–mobility trends. 
    (\textbf{4}) Unsupervised learning is used to classify ions into diffusion modes based on local environments. 
    (\textbf{5}) Human-guided interpretation is used to assign physical meaning to each cluster (e.g., surface diffusion, interlayer migration). 
    (\textbf{6}) Supervised learning and SHAP analysis identify which features govern diffusion class assignment. 
    This combined approach enables interpretable classification and prediction of ion transport in complex amorphous materials.
    }
    \label{fig:workflow_new}
\end{figure}

We begin by outlining our integrated computational framework (Fig. \ref{fig:workflow_new}), designed to uncover sodium storage and transport mechanisms in hard carbon electrodes with diverse nanodomains and heterogeneous microstructures. Although this approach is broadly applicable to various carbon architectures, we focus here on highly disordered carbon to highlight the methodology’s strengths. Disordered carbon’s structural and compositional complexity makes it an ideal system for probing subtle structure–transport relationships. Our framework combines large-scale molecular dynamics (MD) simulations powered by machine-learning interatomic potentials, physically informed structural descriptors, and data-driven feature extraction to classify and interpret a wide range of diffusion behaviors.

At the core of the framework are MD simulations (Fig. \ref{fig:workflow_new}, Panel 1) using the Allegro machine-learned interatomic potential (MLIP)~\cite{musaelian2023learning, kozinsky2023scaling, tan2025high}, which is trained on DFT data for crystalline graphene, amorphous carbon, and sodium-intercalated structures (see SI Section 1.1 for details). This enables simulations that are both accurate and scalable. We systematically vary carbon density (1.5–3.0 g/cc) and sodium loading (25\%, 50\%, and 100\%) to explore how global structure and ion concentration affect transport kinetics (see SI Section 1.2 for details). While experimental bulk densities for hard carbon typically range from 1.5 to 1.9 g/cc~\cite{zhang2016correlation, hyun2024design, irisarri2015hard}, our simulations extend up to 3.0 g/cc to capture both experimentally relevant and idealized motifs. This broader range allows us to sample local environments, such as densely packed or compressed regions, that may arise during synthesis. By including these higher-density configurations, we capture a more diverse spectrum of microstructures and systematically assess how design parameters like carbon density and sodium loading influence microstructural evolution and ion transport. Notably, the significant atomic-level heterogeneity present in disordered carbon, even at fixed density and capacity, renders traditional average-based analyses inadequate.

To resolve atomic-scale heterogeneity, we perform per-ion, per-frame structural analysis (Fig. \ref{fig:workflow_new}, Panel 2). For each sodium ion, we extract four physically motivated descriptors that characterize its local environment: (i) Na–Na coordination, quantifying ionic crowding and clustering; (ii) Na–C coordination, reflecting binding to the carbon framework; (iii) tortuosity, measuring the complexity of local diffusion pathways; and (iv) accessible volume, representing the physical space available for ion motion (see SI Section 1.4 for details). These descriptors are selected for their strong theoretical and empirical connections to ion mobility in amorphous and confined systems. Importantly, they provide interpretable proxies for microstructure without requiring manual labeling or domain segmentation.

To link local structure with transport, we compute ion-specific diffusion coefficients and correlate them with these descriptors (Fig. \ref{fig:workflow_new}, Panel 3), generating a continuous mapping between microstructure and transport behavior. This analysis begins to reveal dominant trends, but the relationships between descriptors and diffusion are highly complex, descriptor spaces are broad, nonlinear, and interdependent, making it difficult to define simple mechanistic rules (see SI Section 1.5). Pairwise correlations alone cannot capture the diversity of ion trajectories in disordered systems.

To address this, we employ machine learning techniques capable of uncovering patterns in high-dimensional, entangled feature spaces. Specifically, we apply unsupervised learning to identify and label emergent diffusion modes based on per-ion structural and dynamical descriptors (Fig. \ref{fig:workflow_new}, Panel 4). Agglomerative clustering groups ions into distinct populations with shared structural–dynamical signatures, capturing the coexistence of behaviors such as concerted diffusion, cavity hopping, and trapping within the same system (see SI Section 1.6). These clusters are both data-driven and physically meaningful, aligning with established knowledge of ion motion in confined and disordered environments.

We then classify each ion’s diffusion behavior (Fig. \ref{fig:workflow_new}, Panel 5) by assigning it to mechanistic modes, including surface-driven diffusion, interlayer hopping, clustering-dominated transport, and immobilized states. Human interpretation is crucial at this stage, that is, by inspecting representative trajectories and structural motifs, we align data-derived clusters with physically meaningful diffusion mechanisms. This human-in-the-loop step ensures that the results are grounded in real transport physics and avoids purely statistical overfitting.

Finally, to determine which structural features most strongly govern diffusion mode, we train a supervised XGBoost classifier using the extracted descriptors (Fig. \ref{fig:workflow_new}, Panel 6). Feature attribution via SHAP (SHapley Additive exPlanations) quantifies the influence of Na–Na clustering, tortuosity, and Na–C bonding across diffusion modes, directly linking structural motifs to functional behavior. We further compute Pearson correlations between global design variables (such as carbon density and sodium loading) and the fraction of ions in each transport mode, establishing a high-level processing–structure–property map.

This multistep workflow systematically addresses the complexity of sodium diffusion in hard carbon: atomistic simulations capture raw dynamics; local descriptors resolve heterogeneity; correlation mapping explains continuous trends; clustering extracts mechanistic states; classification grounds these states in physical language; and supervised learning quantifies structure–function relationships. By integrating these elements, our platform delivers a coherent and extensible approach to decoding ion transport physics and guiding the rational design of next-generation hard carbon anodes.

\subsubsection*{Structure–Transport Relationships Across Density and Loading Regimes}

\begin{figure}[H]
    \centering
    \includegraphics[width=\textwidth]{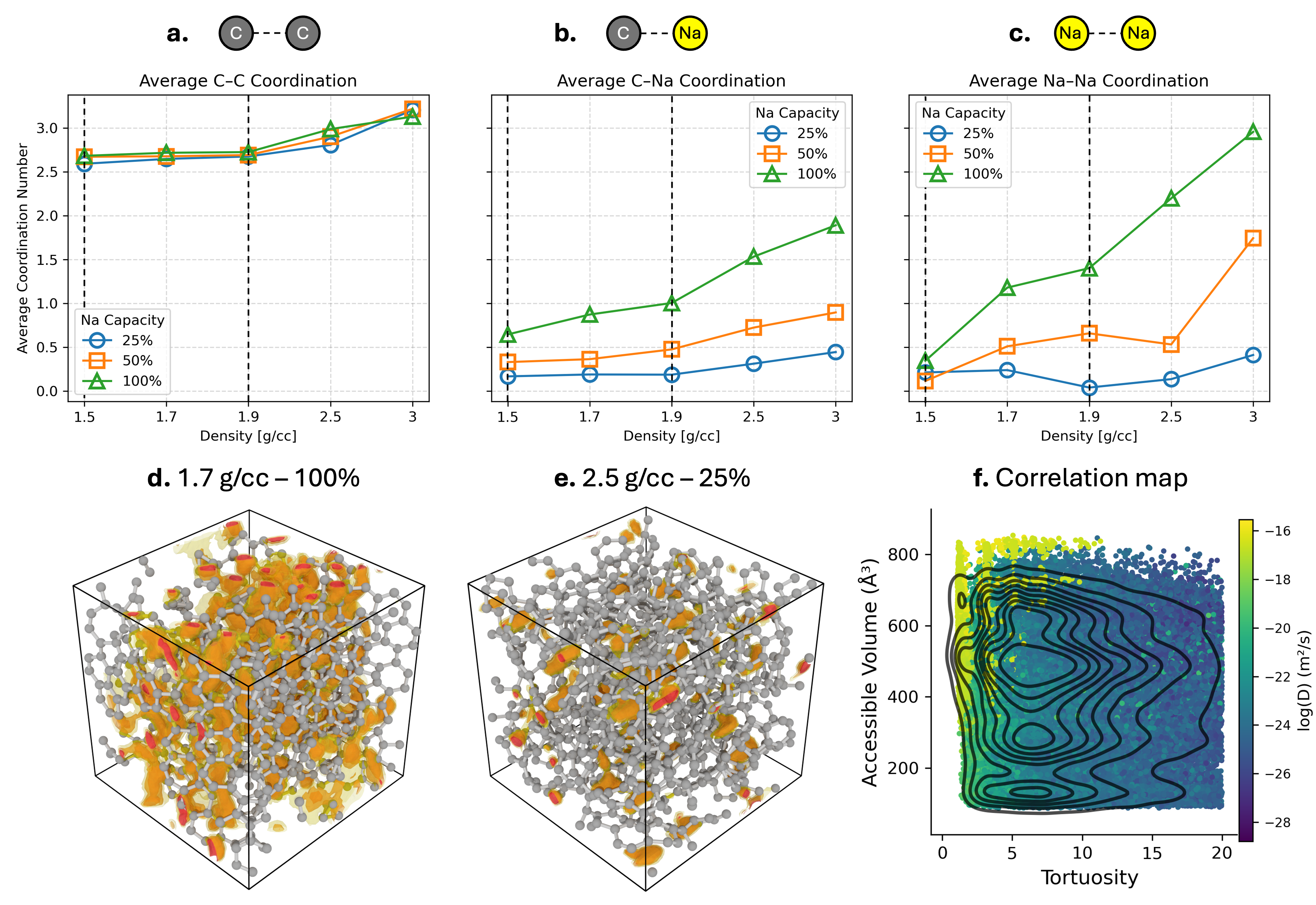}
    \caption{
    \textbf{Structure–transport relationships in Na–C systems across carbon densities and Na-capacities.}
    (\textbf{a–c}) Ensemble-averaged coordination numbers for (\textbf{a}) carbon–carbon (C–C), (\textbf{b}) carbon–sodium (C–Na), and (\textbf{c}) sodium–sodium (Na–Na) interactions across densities (1.5–3.0 g/cc) for 25\%, 50\%, and 100\% sodium loading.
    (\textbf{d, e}) Three-dimensional sodium density isosurfaces (yellow/orange) within gray carbon networks for selected systems.
    (\textbf{f}) Correlation map relating accessible volume and tortuosity to sodium diffusivity ($\log D$), demonstrating key structural controls on transport behavior.
    The vertical dashed lines indicate the typical experimental range of bulk densities observed in hard carbon materials.}
    \label{fig:coordination_sodium}
\end{figure}

With this pipeline in place, we next examine how structural descriptors evolve as a function of carbon density and sodium loading. Figure~\ref{fig:coordination_sodium} provides a comprehensive analysis of these effects. Panels (a–c) present ensemble-averaged coordination numbers (C–C, C–Na, Na–Na), which reveal global structural trends in response to changes in carbon density and sodium content. Panels (d) and (e) display three-dimensional isosurfaces of sodium density, visualizing the degree of ion confinement and connectivity within the carbon matrix. Panel (f) shows two-dimensional histograms that correlate sodium diffusivity with local structural descriptors. Together, these results illustrate how carbon structure and sodium content jointly modulate atomic-scale diffusion by shaping coordination environments, ion distribution, and overall diffusivity. While carbon density determines the compactness of the matrix, significant local heterogeneity often persists, especially in disordered systems, and this heterogeneity strongly influences diffusion behavior.

Figure~\ref{fig:coordination_sodium}a shows that carbon–carbon (C–C) coordination increases modestly with bulk carbon density, as expected, but remains largely insensitive to sodium loading. This finding suggests that the intrinsic topology of the carbon network is robust to compositional changes. In contrast, carbon–sodium (C–Na) coordination, shown in Figure~\ref{fig:coordination_sodium}b, rises with both increasing density and sodium content, reflecting enhanced host–guest interactions as densification brings sodium ions closer to the carbon framework. Sodium–sodium (Na–Na) coordination, presented in Figure~\ref{fig:coordination_sodium}c, exhibits the strongest dependence on both density and loading, increasing substantially at higher values. This trend highlights the onset of ionic crowding, which can promote local clustering or cooperative interactions that restrict mobility through electrostatic repulsion or steric hindrance.

To further elucidate how these coordination patterns affect transport, we visualize three-dimensional sodium density isosurfaces in panels (d) and (e). At a carbon density of 1.7~g/cm$^3$ and 100\% sodium loading, extended and partially interconnected sodium domains emerge within the carbon, enabling moderately percolated transport pathways. In contrast, at 2.5~g/cm$^3$ and 25\% loading, sodium clusters are sparse and spatially confined, indicating more localized and isolated diffusion. These qualitative trends are quantitatively supported by Figure~\ref{fig:coordination_sodium}f, which presents a heatmap correlating sodium diffusivity with key local descriptors. High mobility is associated with low tortuosity and large accessible volume, characteristic of open and navigable pore structures. Conversely, increased Na–Na coordination correlates negatively with diffusion, consistent with immobilization arising from ionic crowding or clustering.

Several key conclusions emerge from this analysis. First, the structure of the carbon matrix varies significantly with density, underscoring the influence of processing conditions. Densification alters both local coordination environments and the overall pore topology, as evidenced by changes in pore size distribution and network connectivity (see SI Section 1.3). Second, sodium–carbon and sodium–sodium interactions both intensify with increasing density and sodium content, leading to greater binding and clustering effects. Most importantly, ion transport is governed by the local structural environment: diffusion is favored in regions characterized by low tortuosity, moderate coordination, and high accessible volume. Collectively, these findings indicate that optimal sodium mobility is achieved at intermediate densities and loadings, where the balance between confinement and percolation maximizes transport.


\subsubsection*{Unsupervised Discovery of Sodium-Ion Diffusion Modes}

\begin{figure}[H]
    \centering
    \includegraphics[width=\textwidth]{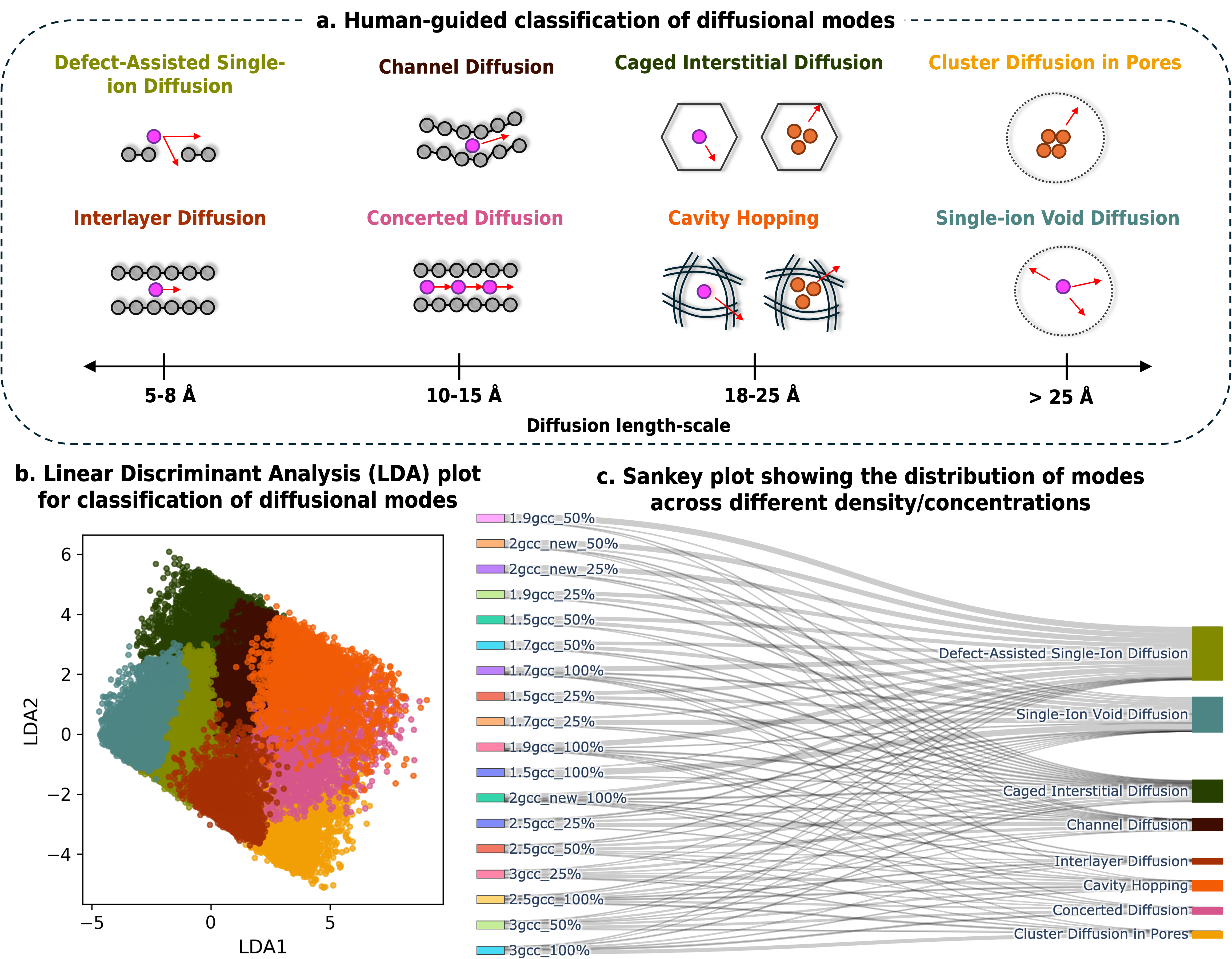}
    \caption{\textbf{Machine learning classification of sodium-ion diffusion modes across structural regimes.}  
    (a) Schematic illustrations of eight distinct Na-ion diffusional mechanisms identified via unsupervised learning, including single-ion and cluster-based transport modes: Caged Interstitial Diffusion, Defect-Assisted Single-Ion Diffusion, Channel Diffusion, Interlayer Diffusion, Cluster Diffusion in Pores, Single-Ion Void Diffusion, Cavity Hopping, and Concerted Diffusion.  
    (b) Linear Discriminant Analysis (LDA) projection of the diffusion modes reveals well-separated clusters, indicating that local descriptors (e.g., coordination, accessible volume, tortuosity) encode sufficient discriminatory information.  
    (c) Sankey diagram showing the prevalence of each diffusion mode across combinations of density (1.5--3.0 g/cc) and sodium loading (25\%--100\%), demonstrating how structural and compositional factors regulate dominant ion transport mechanisms.}
    \label{fig:classification}
    \end{figure}

Building on our understanding of fundamental structure–transport relationships, we next perform a mode-resolved analysis of ion dynamics. Specifically, we classify sodium-ion trajectories into physically interpretable diffusion modes (Fig.~\ref{fig:classification}) by applying unsupervised machine learning to the same set of structural descriptors (see SI Section 1.4 for details). This approach establishes a direct link between local structural environments and transport mechanisms, providing deeper insight into how atomic-scale features dictate diffusion behavior.

Figure~\ref{fig:classification} reveals eight distinct diffusion modes, identified through agglomerative clustering of ion-specific descriptors and diffusivity values. These modes encompass a range of mechanisms, including localized single-ion hopping, defect-assisted motion, void trapping, and more collective behaviors such as concerted migration and cluster diffusion. Each mode is characterized by a unique combination of structural environment and ion migration pathway, highlighting the diversity of sodium-ion motion within disordered carbon matrices. The effectiveness of our descriptor set, Na–Na coordination, Na–C coordination, tortuosity, and accessible volume, is further demonstrated in Fig.~\ref{fig:classification}b. Here, Linear Discriminant Analysis (LDA) shows clear separation between diffusion modes in the reduced descriptor space, confirming that these physically meaningful features can robustly distinguish different transport mechanisms. Notably, the assignment of physically meaningful labels to each cluster required human interpretation, underscoring the importance of domain expertise in translating algorithmic results into actionable design principles.

Figure~\ref{fig:classification}c illustrates how the prevalence of each diffusion mode varies with system conditions, as visualized in a Sankey diagram. In low-density and low-capacity systems (e.g., 1.5 g/cm$^3$ and 25\% Na), diffusion is dominated by caged or defect-assisted modes, reflecting limited percolation and strong surface interactions. Intermediate regimes (e.g., 1.7–1.9 g/cm$^3$) favor more mobile transport modes, such as cavity hopping and concerted diffusion, likely due to improved pore connectivity and balanced coordination environments. At the highest densities and sodium loadings (e.g., 3.0 g/cm$^3$ at 100\% Na), cluster formation and void trapping become prevalent, driven by increased crowding and confinement effects.

This mode-resolved classification highlights several key insights. First, sodium-ion transport is intrinsically linked to local structural environments, with each diffusion mode arising from specific configurations of the carbon framework and sodium distribution. Second, unsupervised clustering using physically meaningful descriptors offers a robust, data-driven approach for identifying and labeling diffusion behaviors without the need for prior supervision. Third, the combined effects of carbon density and sodium capacity systematically influence diffusion regimes: low values promote isolated or surface-bound motion, intermediate conditions enable cooperative dynamics, and high values result in collective or arrested states. Importantly, this classification approach extends beyond disordered carbon, providing a generalizable framework for investigating ion transport in other compositionally and structurally complex materials.


\subsubsection*{Mapping Diffusion Mode Prevalence Across Structural Regimes}

\begin{figure}[H]
    \centering
    \includegraphics[width=\textwidth]{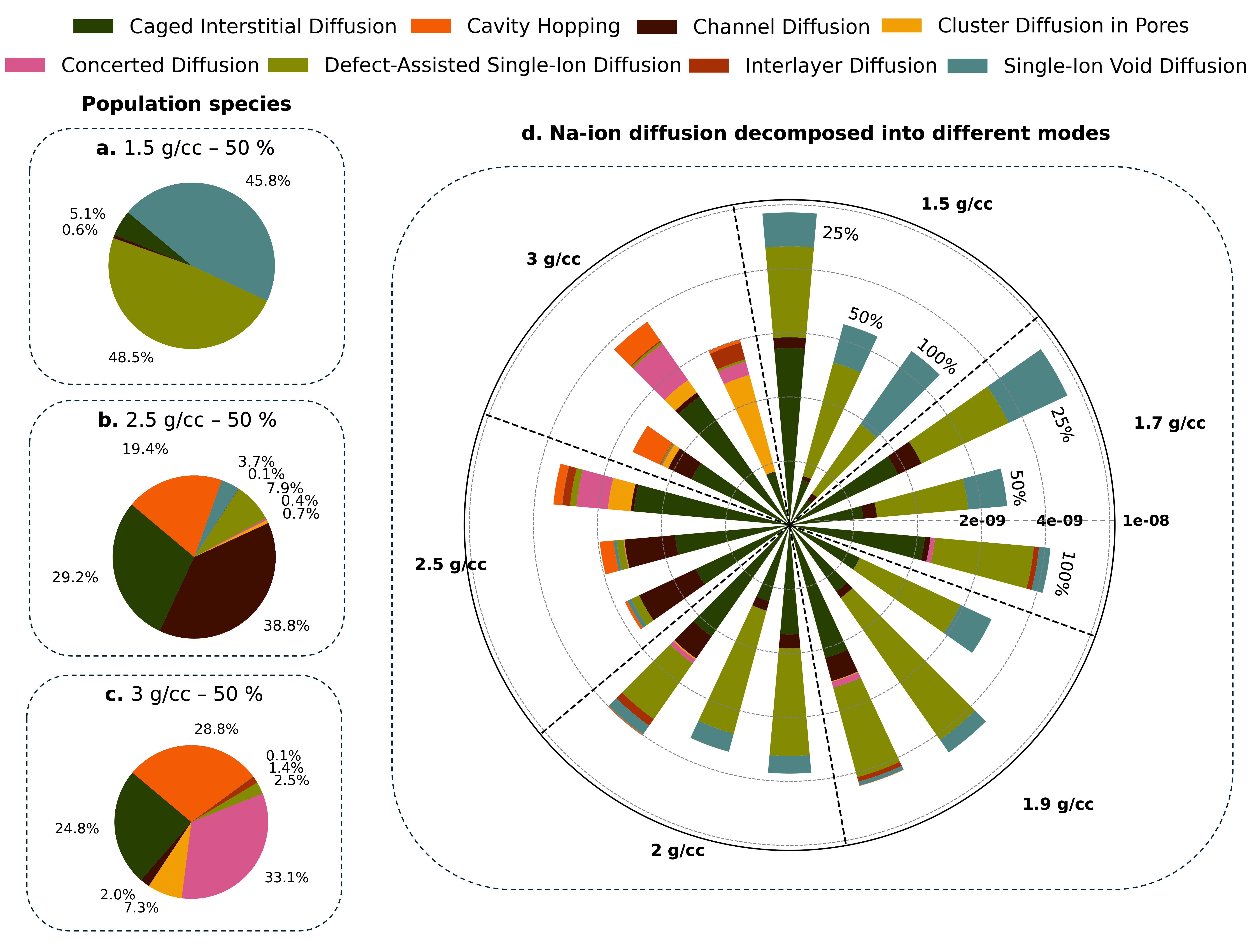}
    \caption{
    \textbf{Sodium diffusion mode decomposition across densities and Na-capacities.} 
    (\textbf{a–c}) Pie charts showing the population distribution of diffusion modes for representative systems at 50\% Na-capacity and densities of 1.5, 2.5, and 3.0 g/cc.  
    (\textbf{d}) Radial bar plot displaying the quantitative breakdown of Na-ion diffusion mechanisms across all density and capacity combinations. Each bar segment represents the fraction of Na-ions exhibiting a specific diffusion mode, color-coded according to the legend. The magnitude reflects the average diffusion coefficient (log scale), revealing the interplay between structure, composition, and transport behavior.
    }
    \label{fig:diffusion_modes}
\end{figure}

After identifying and interpreting the distinct sodium-ion diffusion modes in our systems, we next quantify how their prevalence and transport contributions change across the design space. Through statistical analysis, we correlate the populations and average diffusivities of each diffusion mode with carbon density and sodium loading, as illustrated in Fig.~\ref{fig:diffusion_modes}. This approach offers a comprehensive overview of how various structural configurations promote or inhibit specific ion transport mechanisms.


At low carbon densities (e.g., 1.5 g/cc), sodium-ion diffusion is dominated by single-ion hopping and interlayer transport modes. These systems offer high accessible volume and low tortuosity, enabling relatively unhindered motion with minimal trapping. As carbon density increases to intermediate levels (1.7–1.9 g/cc), we observe a transition toward more cooperative diffusion. Concerted and cluster-based motion become increasingly prevalent, supported by moderately confined yet interconnected channels that facilitate correlated ionic movement.

In contrast, high-density systems (2.5–3.0 g/cc) display a marked shift in diffusion behavior. Here, void diffusion, cavity hopping, and trapped states become more common, reflecting environments with reduced free volume, greater confinement, and increased ionic crowding. Sodium loading further modulates these trends. At fixed density, higher Na-loading amplifies both concerted and cluster diffusion due to stronger inter-ion interactions, but also raises the likelihood of trapping as Na–Na crowding intensifies. This trade-off is visually summarized in the radial bar plot in Fig.~\ref{fig:diffusion_modes}d, which depicts the distribution of diffusion modes across all 18 density–capacity combinations. Lower-density, low-Na systems are enriched in green and blue segments, corresponding to caged interstitial, defect assisted ion and single-ion void diffusion, while high-density, high-capacity conditions show increased orange, magenta, and yellow segments, representing concerted, cavity hopping, and clustered diffusion in pores, respectively.

These findings reinforce several key insights. First, sodium transport behavior is directly determined by local structural environments, whether ions diffuse independently, as clusters, or remain immobilized depends on coordination and available free volume. Second, transport modes systematically evolve with both density and capacity: low-density systems favor fast, independent motion; intermediate regimes enable cooperative dynamics; and high-density or high-capacity states suppress mobility through crowding and confinement. Finally, there is an inherent trade-off between maximizing storage and maintaining high mobility. While increasing Na-content enhances the diversity of diffusion mechanisms, it also introduces bottlenecks through void filling and inter-ionic blocking. These results underscore the importance of carefully tuning both structure and composition to optimize the balance between storage capacity and ionic conductivity.

\subsubsection*{Correlation Mapping of Diffusion Modes and Structural Regimes}

\begin{figure}[H]
    \centering
    \includegraphics[width=\textwidth]{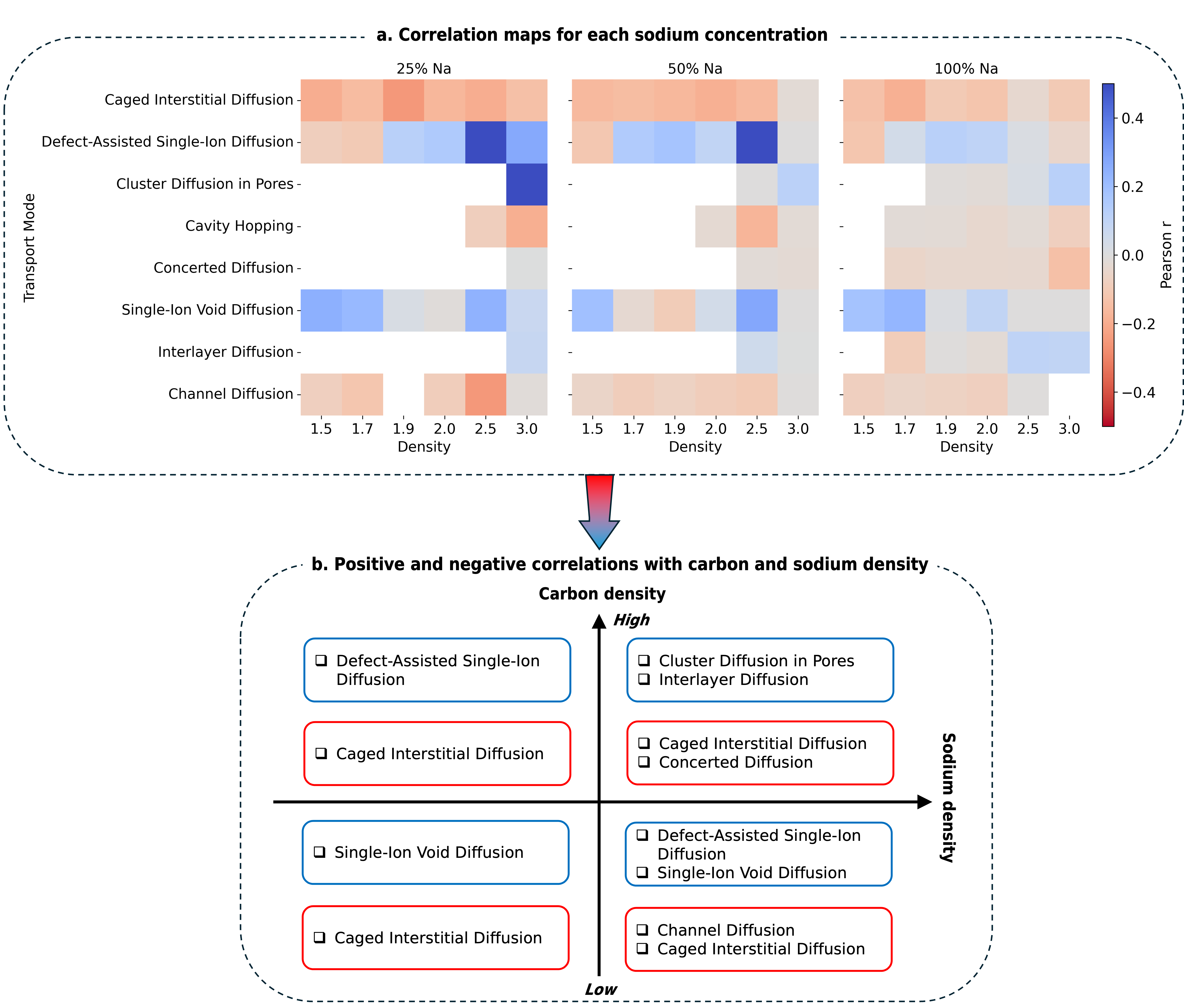}
    \caption{
    \textbf{Correlation between sodium diffusion modes and structural parameters across Na-capacities.}  
    (\textbf{a}) Pearson correlation coefficients ($r$) between the fractional population of each identified diffusion mode and carbon density are shown for systems at 25\%, 50\%, and 100\% Na-capacity. Positive correlations (red) indicate that a transport mode is favored in denser structures, while negative correlations (blue) suggest suppression under those conditions.  
    (\textbf{b}) Schematic quadrant plot summarizing the dominant positive (blue) and negative (red) correlations of each diffusion mode with respect to carbon density and sodium capacity.  
    Together, these panels reveal how distinct transport mechanisms are structurally encoded and selectively activated across density–capacity regime highlighting the influence of crowding, confinement, and network topology on sodium mobility.
    }
    \label{fig:pearson_corr_split}
\end{figure}

Building on this mode-resolved landscape, we quantitatively link diffusion behavior to underlying structural design parameters. To systematically assess the influence of carbon density and sodium loading on specific diffusion modes, we computed Pearson correlation coefficients between the fractional population of each mode and the microscopic control parameters (Fig.~\ref{fig:pearson_corr_split}). This analysis reveals how distinct transport mechanisms are either favored or suppressed under varying structural constraints.

Distinct diffusion modes exhibit clear correlations with density and sodium loading, highlighting regime-specific preferences. Defect-assisted and void diffusion modes show strong positive correlations with high carbon density and sodium loading (e.g., 3.0 g/cc, 100\%), suggesting that ionic crowding and structural disorder promote localized, immobilized behavior. In contrast, caged and channel diffusion are negatively correlated with increasing density and sodium content, indicating that these more mobile transport modes are suppressed under conditions of crowding or confinement. Concerted diffusion and cavity hopping display limited or inconsistent correlations, reflecting their emergence only within narrow or intermediate structural regimes. Cluster diffusion in pores and interlayer diffusion show neutral trends, implying their activation depends more on local geometry than on global density or capacity. These contrasting trends underscore the structural selectivity of different transport mechanisms.

A more detailed breakdown by sodium loading further illustrates how diffusion behavior evolves across concentration regimes. At low Na-loading (25\%), single-ion void diffusion and defect-assisted diffusion are positively correlated with low-to-moderate carbon densities, where open networks and undercoordinated environments facilitate individual ion transport. Cluster diffusion appears only at the highest density, signaling early crowding effects even at low sodium content. Caged and channel diffusion are largely absent or negatively correlated, consistent with the lack of confinement in these open systems. At intermediate loadings (50\%), defect-assisted diffusion is further enhanced under higher carbon densities, while cluster and void diffusion show weak positive correlations with density, marking the onset of collective dynamics. Cavity hopping exhibits a mild negative correlation near intermediate densities, suggesting competition between cooperative and independent ion motion, while caged diffusion remains generally suppressed. At full loading (100\%), ion transport is strongly dictated by crowding. Both defect-assisted and cluster diffusion show strong positive correlations with sodium loading and carbon density, consistent with increased Na–Na coordination and reduced free volume. Caged and hopping mechanisms are largely suppressed, while interlayer and concerted diffusion display neutral or weak correlations, indicating their activation is more stochastic or geometry-dependent under dense loading conditions.

These trends reveal several key insights. First, diffusion modes are highly regime-specific: low Na-loading and low carbon density favor isolated and defect-mediated transport, while high Na-loading and dense carbon structures promote crowding-induced clustering and trapping. Second, sodium concentration serves as an effective control parameter, that is, low loadings enable high mobility within open frameworks, whereas high loadings result in structural saturation and reduced diffusivity. Third, intermediate regimes act as transition points, where multiple transport mechanisms coexist and compete. Finally, this mode-resolved correlation map identifies structural ``design levers'' for optimizing ion transport. By strategically tuning carbon density and sodium loading, it is possible to enhance desirable diffusion modes and minimize trapping, supporting the informed design of carbon-based electrode materials with improved ionic conductivity.


\section*{Conclusion and Discussion}

This work presents a human-guided, physics-informed machine learning framework for dissecting sodium-ion transport in disordered carbon materials, revealing how both structure and ion concentration govern atomic-scale mobility. By integrating large-scale molecular dynamics simulations with physically meaningful, per-ion descriptors and interpretable learning techniques, we bridge the gap between microscopic dynamics and macroscopically tunable processing conditions. This approach delivers a comprehensive, physically grounded understanding of sodium-ion transport within disordered carbon frameworks.

Our analysis shows that increasing carbon density subtly densifies the carbon matrix, resulting in higher C–C coordination, while both Na–C and Na–Na interactions intensify with increasing carbon density and sodium loading. These trends indicate the onset of crowding and confinement effects that shape ion diffusion. However, global averages alone cannot capture the spatial and dynamical heterogeneity inherent in highly disordered systems. To address this, we introduced trajectory-resolved descriptors, such as Na-Na and Na-C coordination, tortuosity, and accessible volume, calculated per ion and per frame. This enabled us to differentiate local structural environments and resolve their impact on ion diffusion. By combining machine-learned interatomic potentials, unsupervised clustering, and supervised correlation analysis, we uncover a spectrum of diffusion modes, ranging from single-ion hopping and interlayer migration to concerted motion, cluster-based transport, and void trapping. The prevalence of these modes varies systematically across structural regimes: low-density systems support mobile, unconfined diffusion; intermediate conditions enable cooperative dynamics; and high carbon and sodium loadings favor trapped or immobilized transport due to pore constriction and crowding. Correlation analysis further links these modes to system-level variables, revealing that mobile diffusion dominates at low carbon density and sodium loading, concerted modes peak at intermediate regimes, and void- or defect-driven modes prevail under extreme crowding.

Overall, our findings establish that sodium-ion transport in disordered carbon structures is neither homogeneous nor random, but is strongly dictated by both local structural motifs and global design parameters. Sodium mobility arises from a complex interplay of local geometric constraints, crowding effects, and network connectivity, rather than being determined solely by global material densities. The interpretable, data-driven framework introduced here, rooted in physical descriptors and guided by human insight, provides a transferable platform for studying ion dynamics in complex disordered systems. Its generality enables future applications to ion and defect transport in porous or amorphous hosts, supporting the rational design of next-generation energy materials.

\section*{Methods}

Disordered carbon structures were generated using a melt–quench molecular dynamics (MD) protocol implemented in LAMMPS~\cite{plimpton1995fast, LAMMPS}, with the Allegro formalism for machine learning interatomic potentials (ML-IAPs)~\cite{musaelian2023learning, kozinsky2023scaling, tan2025high}. Initial atomic configurations were constructed at target carbon densities ranging from 1.5 to 3.0~g/cm$^3$ using the PACKMOL package~\cite{martinez2009packmol}. Each system was equilibrated at 300~K, then heated to 14,000~K for 5~ns to remove structural memory. After a 1~ns hold at this temperature, the system was linearly quenched to 1000~K and further equilibrated for 10~ns in the NVT ensemble using a Nosé–Hoover thermostat~\cite{nose1984unified, hoover1985canonical}.

Sodium ions were inserted into the equilibrated carbon matrices at 25\%, 50\%, and 100\% loadings to represent different sodiation levels. All simulations were performed in a cubic box with a length of 20~\AA. The ML-IAP was trained on a dataset of density functional theory (DFT) reference calculations, spanning crystalline, amorphous, and defective Na–C configurations across a range of temperatures and sodium concentrations. Model validation against \textit{ab initio} molecular dynamics (AIMD) simulations at 800~K and 1200~K showed strong agreement in radial distribution functions (RDFs) and vibrational density of states, confirming the potential’s accuracy for both structural and dynamical properties.

Trajectory-resolved structural analysis was conducted to characterize the local environment of each sodium ion. Extracted descriptors included Na–Na and Na–C coordination numbers, accessible volume (AV), and tortuosity. These per-ion features, along with individual diffusion coefficients (computed via the Einstein relation), served as inputs for unsupervised machine learning. Agglomerative hierarchical clustering was used to classify sodium ions into distinct transport modes, enabling a data-driven interpretation of ionic motion without prior labeling. To identify key structural factors governing diffusion, a supervised XGBoost~\cite{chen2016xgboost} classifier was trained on the same feature set. Feature importance was assessed using SHAP (SHapley Additive exPlanations) values~\cite{lundberg2017unified}, which quantify the marginal contribution of each descriptor to the model’s output across all transport modes. Pearson correlation analysis between system-level parameters (density and sodium loading) and SHAP-derived feature contributions yielded a quantitative structure–function map of sodium ion dynamics. Additional algorithmic details, hyperparameters, and validation analyses are provided in the Supplementary Information.

\section*{Conflicts of interest}
There are no conflicts to declare.

\section*{Acknowledgments}
This work was performed under the auspices of the U.S. Department of Energy by Lawrence Livermore National Laboratory under contract DE-AC52-07NA27344. This material is based upon work supported by the U.S. Department of Energy, Office of Electricity (OE), Energy Storage Division. Computing support was provided by the Lawrence Livermore National Laboratory Institutional Computing Grand Challenge program and resources sponsored by the Department of Energy’s Office of Energy Efficiency and Renewable Energy, located at the National Renewable Energy Laboratory.

\printbibliography

@Article{LAMMPS,
  author = "A. P. Thompson and H. M. Aktulga and R. Berger and 
     D. S. Bolintineanu and W. M. Brown and P. S. Crozier and
     P. J. in 't Veld and A. Kohlmeyer and S. G. Moore and T. D. Nguyen and
     R. Shan and M. J. Stevens and J. Tranchida and C. Trott and S. J. Plimpton",
  title = "{LAMMPS} - a flexible simulation tool for
     particle-based materials modeling at the 
     atomic, meso, and continuum scales",
  journal = "Comp. Phys. Comm.",
  volume =  "271",
  pages =   "108171",
  year =    "2022",
  doi = "10.1016/j.cpc.2021.108171"
}

@article{martinez2009packmol,
  title={PACKMOL: A package for building initial configurations for molecular dynamics simulations},
  author={Mart{\'\i}nez, Leandro and Andrade, Ricardo and Birgin, Ernesto G and Mart{\'\i}nez, Jos{\'e} Mario},
  journal={Journal of computational chemistry},
  volume={30},
  number={13},
  pages={2157--2164},
  year={2009},
  publisher={Wiley Online Library}
}

@article{li2022insight,
  title={Insight into sodium storage behaviors in hard carbon by reaxff molecular dynamics simulation},
  author={Li, Jiaqi and Peng, Chen and Li, Jie and Wang, Jingkun and Zhang, Hongliang},
  journal={Energy \& Fuels},
  volume={36},
  number={11},
  pages={5937--5952},
  year={2022},
  publisher={ACS Publications}
}

@article{sun2022understanding,
  title={Understanding of sodium storage mechanism in hard carbons: ongoing development under debate},
  author={Sun, Ning and Qiu, Jieshan and Xu, Bin},
  journal={Advanced Energy Materials},
  volume={12},
  number={27},
  pages={2200715},
  year={2022},
  publisher={Wiley Online Library}
}

@article{fan2024research,
  title={Research progress on hard carbon materials in advanced sodium-ion batteries},
  author={Fan, Xiangyu and Kong, Xirui and Zhang, Pengtang and Wang, Jiulin},
  journal={Energy Storage Materials},
  volume={69},
  pages={103386},
  year={2024},
  publisher={Elsevier}
}

@article{chen2021hard,
  title={Hard carbon for sodium storage: mechanism and optimization strategies toward commercialization},
  author={Chen, Dequan and Zhang, Wen and Luo, Kangying and Song, Yang and Zhong, Yanjun and Liu, Yuxia and Wang, Gongke and Zhong, Benhe and Wu, Zhenguo and Guo, Xiaodong},
  journal={Energy \& Environmental Science},
  volume={14},
  number={4},
  pages={2244--2262},
  year={2021},
  publisher={Royal Society of Chemistry}
}

@article{chu2023reconfiguring,
  title={Reconfiguring hard carbons with emerging sodium-ion batteries: a perspective},
  author={Chu, Yue and Zhang, Jun and Zhang, Yibo and Li, Qi and Jia, Yiran and Dong, Ximan and Xiao, Jing and Tao, Ying and Yang, Quan-Hong},
  journal={Advanced Materials},
  volume={35},
  number={31},
  pages={2212186},
  year={2023},
  publisher={Wiley Online Library}
}

@article{xie2020hard,
  title={Hard carbons for sodium-ion batteries and beyond},
  author={Xie, Fei and Xu, Zhen and Guo, Zhenyu and Titirici, Maria-Magdalena},
  journal={Progress in Energy},
  volume={2},
  number={4},
  pages={042002},
  year={2020},
  publisher={IOP Publishing}
}

@article{xiao2025insight,
  title={Insight into the Role of Closed-Pore Size on Rate Capability of Hard Carbon for Fast-Charging Sodium-Ion Batteries},
  author={Xiao, Shuhao and Guo, Yu-Jie and Chen, Han-Xian and Liu, Haizhou and Lei, Zhou-Quan and Huang, Lin-Bo and Jin, Ruo-Xi and Su, Xiao-Chuan and Zhang, Qianyu and Guo, Yu-Guo},
  journal={Advanced Materials},
  pages={2501434},
  year={2025},
  publisher={Wiley Online Library}
}

@article{li2022towards,
  title={Towards an atomistic understanding of hard carbon electrode materials and sodium behaviors},
  author={Li, Jiaqi and Peng, Chen and Wang, Jingkun and Li, Jie and Zhang, Hongliang},
  journal={Diamond and Related Materials},
  volume={129},
  pages={109355},
  year={2022},
  publisher={Elsevier}
}

@article{niu2025sodium,
  title={Sodium cluster-driven safety concerns of sodium-ion batteries},
  author={Niu, Jiaping and Dong, Junyuan and Zhang, Xiaohu and Huang, Lang and Lu, Guoli and Han, Xiaolei and Wang, Jinzhi and Gong, Tianyu and Chen, Zheng and Zhao, Jingwen and others},
  journal={Energy \& Environmental Science},
  year={2025},
  publisher={Royal Society of Chemistry}
}

@article{anji2018insight,
  title={Insight into sodium insertion and the storage mechanism in hard carbon},
  author={Anji Reddy, M and Helen, M and Gro{\ss}, Axel and Fichtner, Maximilian and Euchner, Holger},
  journal={ACS Energy Letters},
  volume={3},
  number={12},
  pages={2851--2857},
  year={2018},
  publisher={ACS Publications}
}

@article{musaelian2023learning,
  title={Learning local equivariant representations for large-scale atomistic dynamics},
  author={Musaelian, Albert and Batzner, Simon and Johansson, Anders and Sun, Lixin and Owen, Cameron J and Kornbluth, Mordechai and Kozinsky, Boris},
  journal={Nature Communications},
  volume={14},
  number={1},
  pages={579},
  year={2023},
  publisher={Nature Publishing Group UK London}
}

@article{tan2025high,
  title={High-performance training and inference for deep equivariant interatomic potentials},
  author={Tan, Chuin Wei and Descoteaux, Marc L and Kotak, Mit and Nascimento, Gabriel de Miranda and Kavanagh, Se{\'a}n R and Zichi, Laura and Wang, Menghang and Saluja, Aadit and Hu, Yizhong R and Smidt, Tess and others},
  journal={arXiv preprint arXiv:2504.16068},
  year={2025}
}

@inproceedings{kozinsky2023scaling,
  title={Scaling the leading accuracy of deep equivariant models to biomolecular simulations of realistic size},
  author={Kozinsky, Boris and Musaelian, Albert and Johansson, Anders and Batzner, Simon},
  booktitle={Proceedings of the International Conference for High Performance Computing, Networking, Storage and Analysis},
  pages={1--12},
  year={2023}
}

@article{plimpton1995fast,
  title={Fast parallel algorithms for short-range molecular dynamics},
  author={Plimpton, Steve},
  journal={Journal of Computational Physics},
  volume={117},
  number={1},
  pages={1--19},
  year={1995},
  publisher={Elsevier}
}

@article{nose1984unified,
  title={A unified formulation of the constant temperature molecular dynamics methods},
  author={Nosé, Shuichi},
  journal={The Journal of Chemical Physics},
  volume={81},
  number={1},
  pages={511--519},
  year={1984},
  publisher={AIP Publishing}
}

@article{hoover1985canonical,
  title={Canonical dynamics: Equilibrium phase-space distributions},
  author={Hoover, William G},
  journal={Physical Review A},
  volume={31},
  number={3},
  pages={1695},
  year={1985},
  publisher={APS}
}

@article{lundberg2017unified,
  title={A unified approach to interpreting model predictions},
  author={Lundberg, Scott M and Lee, Su-In},
  journal={Advances in Neural Information Processing Systems},
  volume={30},
  pages={4765--4774},
  year={2017}
}

@article{zhang2016correlation,
  title={Correlation between microstructure and Na storage behavior in hard carbon},
  author={Zhang, Biao and Ghimbeu, Cam{\'e}lia Matei and Laberty, Christel and Vix-Guterl, Cathie and Tarascon, Jean-Marie},
  journal={Advanced Energy Materials},
  volume={6},
  number={1},
  pages={1501588},
  year={2016},
  publisher={Wiley Online Library}
}

@article{kitsu2025microstructure,
  title={Microstructure-Dependent Sodium Storage Mechanisms in Hard Carbon Anodes},
  author={Kitsu Iglesias, Luis and Marks, Samuel D and Rampal, Nikhil and Antonio, Emma N and de Ferreira de Menezes, Rafael and Zhang, Liang and Olds, Daniel and Weitzner, Stephen E and Sprenger, Kayla G and Wan, Liwen F and others},
  journal={Small},
  pages={2505561},
  year={2025},
  publisher={Wiley Online Library}
}

@article{hyun2024design,
  title={Design guidelines for a high-performance hard carbon anode in sodium ion batteries},
  author={Hyun, Jong Chan and Jin, Hyeong Min and Kwak, Jin Hwan and Ha, Son and Kang, Dong Hyuk and Kim, Hyun Soo and Kim, Sion and Park, Minhyuck and Kim, Chan Yeol and Yoon, Juhee and others},
  journal={Energy \& Environmental Science},
  volume={17},
  number={8},
  pages={2856--2863},
  year={2024},
  publisher={Royal Society of Chemistry}
}

@article{irisarri2015hard,
  title={Hard carbon negative electrode materials for sodium-ion batteries},
  author={Irisarri, E and Ponrouch, A and Palacin, MR},
  journal={Journal of The Electrochemical Society},
  volume={162},
  number={14},
  pages={A2476},
  year={2015},
  publisher={IOP Publishing}
}

@inproceedings{chen2016xgboost,
  title     = {XGBoost: A Scalable Tree Boosting System},
  author    = {Chen, Tianqi and Guestrin, Carlos},
  booktitle = {Proceedings of the 22nd ACM SIGKDD International Conference on Knowledge Discovery and Data Mining},
  pages     = {785--794},
  year      = {2016},
  organization = {ACM},
  doi       = {10.1145/2939672.2939785}
}

\end{document}